\documentclass[conference]{IEEEtran}
\IEEEoverridecommandlockouts
\usepackage{cite}
\usepackage{amsmath,amssymb,amsfonts}
\usepackage{graphicx}
\usepackage{textcomp}
\usepackage{xcolor}
\usepackage{dsfont}
\usepackage{bm}

\usepackage{algorithm2e}
\RestyleAlgo{ruled}
\newcommand\norm[1]{\left\lVert#1\right\rVert}

\newcommand\A{\mathbf{A}}
\newcommand\B{\mathbf{B}}
\newcommand{\C}{\mathbf{C}}
\newcommand\D{\mathbf{D}}
\newcommand\I{\mathbf{I}}

\def\BibTeX{{\rm B\kern-.05em{\sc i\kern-.025em b}\kern-.08em
    T\kern-.1667em\lower.7ex\hbox{E}\kern-.125emX}}
    
\usepackage{hyperref}

\usepackage{fancyhdr}
\usepackage{kantlipsum}
\fancypagestyle{plain}{
\fancyhf{}
\fancyhead[C]{Conference on \LaTeX} 

}
\usepackage{eso-pic}
\begin{document}

\AddToShipoutPictureBG*{
\AtPageUpperLeft{
\setlength\unitlength{1in}
\hspace*{\dimexpr0.5\paperwidth\relax}
\makebox(0,-0.75)[c]{\textbf{2022 IEEE/ACM International Conference on Advances in Social Networks Analysis and Mining (ASONAM)}}}}

\title{A Fast Algorithm for Ranking Users by their Influence in Online Social Platforms}

\author{\IEEEauthorblockN{Nouamane Arhachoui, Esteban Bautista, Maximilien Danisch, Anastasios Giovanidis}
\IEEEauthorblockA{\textit{ Sorbonne Universit\'e, CNRS -- LIP6, F-75005 Paris, France}\\
{\{nouamane.arhachoui,\ esteban.bautista-ruiz,\ anastasios.giovanidis\}@lip6.fr}
}
}


\maketitle

\IEEEoverridecommandlockouts
\IEEEpubid{\parbox{\columnwidth}{\vspace{8pt}
\makebox[\columnwidth][t]{IEEE/ACM ASONAM 2022, November 10-13, 2022}
\makebox[\columnwidth][t]{978-1-6654-5661-6/22/\$31.00~\copyright\space2022 IEEE} \hfill}
\hspace{\columnsep}\makebox[\columnwidth]{}}
\IEEEpubidadjcol
\begin{abstract}
Measuring the influence of users in social networks is key for numerous applications. A recently proposed influence metric, coined as $\psi$-score, allows to go beyond traditional centrality metrics, which only assess structural graph importance, by further incorporating the rich information provided by the posting and re-posting activity of users. The $\psi$-score is shown in fact to generalize PageRank for non-homogeneous node activity. Despite its significance, it scales poorly to large datasets; for a network of $N$ users, it requires to solve $N$ linear systems of equations of size $N$. To address this problem, this work introduces a novel scalable algorithm for the fast approximation of $\psi$-score, named \textit{Power}-$\psi$. The proposed algorithm is based on a novel equation indicating that it suffices to solve one system of equations of size $N$ to compute the $\psi$-score. Then, our algorithm exploits the fact that such a system can be recursively and distributedly approximated to any desired error. This permits the $\psi$-score, summarizing both structural and behavioral information for the nodes, to run as fast as PageRank. We validate the effectiveness of the proposed algorithm, which we release as an open source Python library, on several real-world datasets.

\end{abstract}


\section{Introduction}

\subsection{Context}

Online Social Platforms (OSPs) have become ubiquitous in society. Their proliferation, as well as their massive number of users, has made the quantification of the influence of users in OSPs a task of utmost importance. For example, spotting the so-called influencers of a network is crucial for social scientists to better understand the dynamics that determine new trends \cite{dkempe} or the roots of political polarization \cite{alvim2021multi}. It is also key for companies in order to develop a better marketing of their products \cite{bok2021hot}. Machine learning algorithms are another example, where identifying a reduced set of users whose features are present over the entire network is essential to reduce the number of parameters and to avoid the curse of dimensionality \cite{MustoLGS21, cabannes2021overcoming}. 

Social platforms usually provide a user with a wall and a news-feed. In the wall, users can post messages. In the news-feed, users see messages that other users, called their leaders (or followees), have posted on their wall. A user can then re-post a message from its news-feed into its wall, allowing posts to diffuse through the network \cite{diffusion_networks}. A fundamental problem is then to quantify the influence of users. Namely, the degree to which their posts spread in the network and are visible to others. 

In order to address the aforementioned problem, classical centrality metrics \cite{centrality} permit to assign a ranking to the nodes of the network according to their structural importance. However, such metrics are unsatisfactory to assess influence in OSPs, as they do not consider the posting and sharing activity of users in their scores. The relevance of this information is highlighted in \cite{Cha_Haddadi}, where it is shown that users with the highest number of followers (in-degree) may not necessarily be the ones that generate the highest number of retweets or impressions. This means that the in-degree, which is correlated with the popularity of a user, is not the only factor that determines influence. Similarly, the authors in \cite{Sala12} address the question ``Are social links valid
indicators of real user interaction?" and analyze a large user trace from Facebook to showcase the difference. As a result, it is necessary to go beyond pure structural metrics. 

To overcome this limitation, a new metric, coined as $\psi$-score, has been recently introduced in \cite{psiscore}.

The $\psi$-score results from an Online Social Platform model, also presented in \cite{psiscore}, in which the diffusion of information is taken into account with activity features for each user. 
In the general case where users post and re-post content at a different frequency from each other, the OSP model accurately incorporates this user activity to the structural information related to the graph, thus providing a more meaningful ranking metric of user influence in OSPs. This metric was shown in \cite[Theorem 5]{psiscore} to equal PageRank in the special case of homogeneous activity, where all users create the same amount of posts in a given time interval and share with the same rate within this time interval.

\subsection{Goal and contributions}

The problem we address in this paper is that the $\psi$-score calculation from \cite{psiscore} does not scale to large graphs since it needs the resolution of $N$ linear systems of $N$ equations each, $N$ being the number of users in the network, whereas the PageRank would only require to solve one system with $N$ unknowns.

We thus propose a new algorithm based on an equation that revamps these $N$ system into a single one, still with $N$ equations. Our algorithm, which accepts a recursive and distributed implementation, allows to approximate the $\psi$-score of the OSP with a similar speed as the traditional approximation for PageRank via the power method. Hence, the $\psi$-score can be used as a more expressive alternative to PageRank in the task of ranking users by their influence. 

Finally, we provide an open source software library\footnote{\url{https://github.com/NouamaneA/psi-score}} in order to allow anyone to implement the $\psi$-score in their own projects.

\subsection{Related Works}

Centrality measures have been systematically used in various fields of application of network science such as social networks \cite{social_network}, transportation networks \cite{transportation}, communication networks \cite{communication_network}, biological networks \cite{biology} or even political networks \cite{politics}. The most used metrics are degree, betweenness, closeness, eigenvector centrality, and PageRank, among others \cite{centrality, pagerank}.

These metrics, which are practically used to rank nodes of a network, only consider the network topology. For example, PageRank \cite{pagerank} is derived through a random walk with teleportation on the studied graph, taking at each step uniform decisions about which node to consider next.

There are several existing algorithms such as Push \cite{scalable_pagerank} and the use of Chebyshev polynomials \cite{DBLP:journals/snam/BautistaL22} that exploit the random walk interpretation of PageRank in order to accelerate its calculation. The Push method \cite{scalable_pagerank} loops on the nodes by allowing each of them to update the approximation of its own PageRank value and updates its neighbors' values. The method proposed in \cite{DBLP:journals/snam/BautistaL22} uses the Chebyshev polynomials to approximate the PageRank vector in a more efficient way compared to the known power-method introduced with the metric \cite{pagerank}. Although the literature to accelerate graph eigenvalue-based measures is rich, in this paper we will need to focus in what follows on the specific structure of the $\psi$-score with its current algebraic limitations in order to propose an original way to make it scale as fast as PageRank does. Having established in this paper a power-iteration inspired  algorithm that scales, this opens future perspectives to study other methods inspired by the Push method \cite{scalable_pagerank} or Chebyshev polynomials \cite{DBLP:journals/snam/BautistaL22} to achieve even better performance.

\subsection{Outline of the paper}

The paper is structured as follows. Section \ref{sec:II} presents the recently proposed $\psi$-score and discusses its relation to classical centrality metrics. Section \ref{sec:III} contains our main contribution: a new algorithm for the fast approximation of the $\psi$-score that scales to massive datasets. We show that the algorithm runtime is comparable to PageRank, thus providing a more-diverse alternative. Section \ref{sec:IV} briefly presents our software contribution: an open source Python library for the $\psi$-score computation. Section \ref{sec:V} evaluates the performance of our new algorithm on several real-world networks. Conclusions and future work are discussed in Section \ref{sec:VI}.

\begin{table}[t]
    \caption{Notations used in the paper}
    \begin{center}
    \begin{tabular}{|c|c|}
        \hline
        \textbf{Notation} & \textbf{Description}\\
        \hline
        $\psi_i$ & $\psi$-score of user $i$\\
        \hline
        $\bm{\psi}$ & $\psi$-score column-vector\\
        \hline
        $p_i^{(n)}$ & impressions of $i$ on the Newsfeed of $n$\\
        \hline
        $q_i^{(n)}$ & (influence) impressions of $i$ on the Wall of $n$\\
        \hline
        $\mathbf{p}_i$ & vector with every $p_i^{(n)}$\\
        \hline
        $\mathbf{q}_i$ & vector with every $q_i^{(n)}$\\
        \hline
        $\mathbf{P}$ & matrix  with $\mathbf{p}_i$ as column $i$\\
        \hline
        $\mathbf{Q}$ & matrix with $\mathbf{q}_i$ as column $i$\\
        \hline
        $\mathbf{b}_i$ & input vector of normalised posting activity \\
        \hline
        $\mathbf{d}_i$ & vector with the posting ratio of a user $i$\\
        \hline
        $\mathbf{B}$ & matrix with $\mathbf{b}_i$ as column $i$\\
        \hline
        $\mathbf{D}$ & diagonal matrix with $\mathbf{d}_i$ as column $i$\\
        \hline
        $\mathbf{1}$ & column-vector of $\mathbb{R}^N$ full of ones, i.e. $(1 \; 1 \; \cdots \; 1)^T$\\
        \hline
    \end{tabular}
    \end{center}
    \label{tab:notations}
\end{table}

\section{The $\psi$-score}
\label{sec:II}

\begin{figure}[t]
    \centering
    \includegraphics[width=\linewidth]{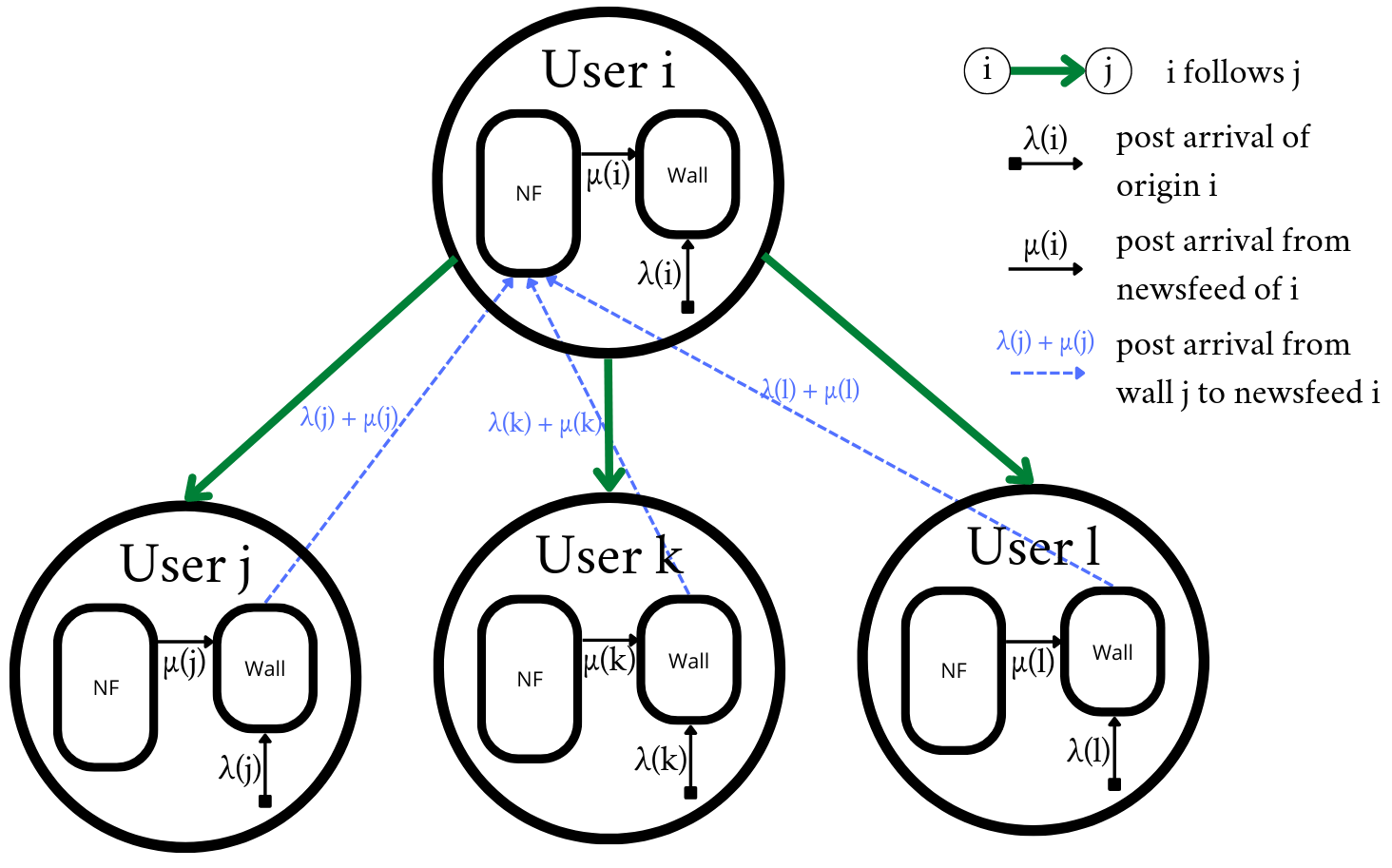}
    \caption{A small example of the Online Social Platform from the point of view of user $i$.}
    \label{fig:osp}
\end{figure}

Recently, a new score of users' influence over their social network (platform) has been introduced in \cite{psiscore}. The calculation of this score involves information not only about the users' positions inside the graph structure, but additionally their posting and sharing activity. These are two extra important features that enrich previous graph-based metrics: when a user posts more frequently he/she should have a higher influence because his/her posts appear more often on the newsfeed of others; furthermore, when a user shares content rarely, very few posts (other than of his/her own creation) pass through to reach his/her followers.

More formally, let $\mathcal{G} = (\mathcal{N}, \mathcal{E})$ be an unweighted and directed graph where $\mathcal{N}$ denotes the set of vertices of cardinality $|\mathcal{N}|=N$ and $\mathcal{E}$ refers to the set of edges of cardinality $|\mathcal{E}| = M$. Vertices model the users of social platforms and edges the follower-leader relations: $(i,j)\in\mathcal{E}$ encodes that user $i$ follows user $j$. Let $\lambda$ be the vector of posting activity over the set of vertices. That is, $\lambda^{(n)}$ is the number of posts per unit of time (posting frequency) that user $n$ creates on his/her wall. Similarly, let $\mu$ be the vector of re-posting/sharing activity over the set of vertices. Then,  $\mu^{(n)}$ refers to the frequency with which user $n$ visits his/her news-feed and chooses one of the current entries uniformly at random to re-post it on his/her wall. 

From now on in this section we will focus on posts of origin $i$. The authors of \cite{psiscore} show that the aforementioned graph and activity features can be used to compute the following two vectors associated with the influence of user $i \in \mathcal{N}$:

\begin{itemize}
    \item $\mathbf{p}_i = (p_i^{(1)} \; p_i^{(2)} \; \cdots\; p_i^{(N)})^T$ where, for $n \in \mathcal{N}$, $p_i^{(n)}$ is the expected percentage of posts originating from user $i$ on the news-feed of user $n$.
    \item $\mathbf{q}_i = (q_i^{(1)} \; q_i^{(2)} \; \cdots\; q_i^{(N)})^T$ where, for all $n \in \mathcal{N}$, $q_i^{(n)}$ is the expected percentage of posts originating from user $i$ on the wall of user $n$.
\end{itemize}

A central point in \cite{psiscore} is that $q_i^{(n)}$ can be interpreted as the influence of user $i$ on user $n$, meaning that the more posts of origin $i$ appear on the wall of $n$, the larger the influence that $i$ has on $n$. This is used to define the influence of user $i$ over the entire network as the average influence of $i$ on all users. This notion is formalized by the so-called $\psi$-score of user $i$ which is defined as 
\begin{equation}
    \label{eq_psiscore}
    \psi_i = \frac{1}{N} \sum\limits_{n\in\mathcal{N}}{q_i^{(n)}}.
\end{equation}

The expression in (\ref{eq_psiscore}) involves all entries of the vector $\mathbf{q}_i$ for the calculation of the influence score $\psi_i$ of user $i$. However, $\mathbf{q}_i$ is not a given information and to find $q_i^{(n)} \; \forall n \in \mathcal{N}$ we need to resolve a system of equations. Specifically, each $q_i^{(n)}$ can be derived from the $p_i^{(n)}$ thanks to the following equations:
\begin{eqnarray}
    \label{balance_eq_wall_1}
    \text{for } n=j\neq i,\; (\lambda^{(j)} + \mu^{(j)}) q_i^{(j)} &=& \mu^{(j)} p_i^{(j)}\\
    \label{balance_eq_wall_2}
    \text{for } n=i,\; (\lambda^{(i)} + \mu^{(i)}) q_i^{(i)} &=& \lambda^{(i)} + \mu^{(i)} p_i^{(i)}
\end{eqnarray}

Moreover, the vector $\mathbf{p}_i$ is the solution of a fixed-point problem (see \cite{psiscore}, section III). For a post of origin $i$ in the Newsfeed of $j$ we have the following balance equation:
\begin{equation}
    \label{balance_eq_newsfeed_1}
    \sum\limits_{k\in \mathcal{L}^{(j)}}(\lambda^{(k)} + \mu^{(k)})p_i^{(j)} = \lambda^{(i)} \mathds{1}_{\{i\in \mathcal{L}^{(j)}\}} + \sum\limits_{k \in \mathcal{L}^{(j)}}\mu^{(k)} p_i^{(k)}
\end{equation}

Since there are no self-loops and a user cannot be its own leader or follower, we have the following balance equation for user $i$'s Newsfeed:
\begin{equation}
    \label{balance_eq_newsfeed_2}
    \sum\limits_{k\in \mathcal{L}^{(i)}}(\lambda^{(k)} + \mu^{(k)})p_i^{(i)} = \sum\limits_{k \in \mathcal{L}^{(i)}}\mu^{(k)} p_i^{(k)}
\end{equation}

The system of equations (\ref{balance_eq_wall_1})-(\ref{balance_eq_wall_2}) and (\ref{balance_eq_newsfeed_1})-(\ref{balance_eq_newsfeed_2}) can be written in matrix form with the following linear system:
\begin{eqnarray}
    \label{newsfeed}
    \mathbf{p}_i &=& \mathbf{A}.\mathbf{p}_i + \mathbf{b}_i\\
    \label{wall}
    \mathbf{q}_i &=& \mathbf{C}.\mathbf{p}_i + \mathbf{d}_i
\end{eqnarray}
where, \\
\begin{math}
    \mathbf{A}\in \mathbb{R}_+^{N\times N}:\; a_{ji} = \frac{\mu^{(i)}}{\sum\limits_{\ell\in \mathcal{L}^{(j)}}(\lambda^{(\ell)}+\mu^{(\ell)})}\mathds{1}_{\{i\in \mathcal{L}^{(j)}\}},\\
    \mathbf{b}_i\in \mathbb{R}_+^{N}:\; b_{ji} = \frac{\lambda^{(i)}}{\sum\limits_{\ell\in \mathcal{L}^{(j)}}(\lambda^{(\ell)}+\mu^{(\ell)})}\mathds{1}_{\{i\in \mathcal{L}^{(j)}\}},\\
    \mathbf{C}\in \mathbb{R}_+^{N\times N}:\; c_{ji} = \frac{\mu^{(j)}}{\lambda^{(j)}+\mu^{(j)}}\mathds{1}_{\{j=i\}},\\
    \mathbf{d}_i\in \mathbb{R}_+^{N}:\; d_{ji} = \frac{\lambda^{(i)}}{\lambda^{(i)}+\mu^{(i)}}\mathds{1}_{\{j=i\}}.\\
\end{math}

To solve the fixed-point problem of (\ref{newsfeed}), we can use the following iteration (see \cite{psiscore}, Theorem 4):
\begin{equation}
    \label{fixpt_algo}
    \mathbf{p}_i(t) = \mathbf{A} \mathbf{p}_i(t-1) + \mathbf{b}_i
\end{equation}
where $t$ is the current iteration. With any initialization of $\mathbf{p}_i(0)$, the $\mathbf{p}_i(t)$ converges towards the solution $\mathbf{p}_i$ of (\ref{newsfeed}) as $t\rightarrow \infty$ because $\mathbf{A}$ is a sub-stochastic matrix. This method is less costly than using matrix inversion to solve the system.\\

\begin{algorithm}[t]
\caption{\texttt{Power-NF} from \cite{psiscore}: Power method for the Newsfeed probabilities $\mathbf{p}_i$ related to user $i$.}
\label{nf_algo}
\SetAlgoLined
\SetKwData{Left}{left}\SetKwData{This}{this}\SetKwData{Up}{up}\SetKwFunction{Union}{Union}\SetKwFunction{FindCompress}{FindCompress}
\SetKwInOut{Input}{input}\SetKwInOut{Output}{output}
\Input{Origin User $i$, $N$ number of users, $N\times N$ matrix $\mathbf{A}$, vector $\mathbf{b}_i$, $\mathbf{p}$-tolerance $\varepsilon$}
\Output{vector $\mathbf{p}_i$}
$\mathbf{p}_i \leftarrow \mathbf{b}_i $\;
$t \leftarrow 0$\;
$gap \leftarrow 1$\;
\While{
$(gap > \varepsilon)$
}
{
    $\mathbf{p}_i^{old} \leftarrow \mathbf{p}_i$\;
    $\mathbf{p}_i \leftarrow \A \mathbf{p}_i^{old} + \mathbf{b}_i$\;
    $gap \leftarrow \norm{\mathbf{p}_i - \mathbf{p}_i^{old}}$\;
    $t \leftarrow t + 1$\;
}
return $\mathbf{p}$\;
\end{algorithm}

To rank all users, we need to calculate $\psi_i$ for all $i\in\mathcal{N}$, i.e. we need the vector $\bm{\psi} = (\psi_1, \; \psi_2, \; \cdots,\; \psi_N)^T \in \mathbb{R}_+^N$. Thus, the iterative algorithm in \eqref{fixpt_algo} has to be applied $N$ times to solve the linear system \eqref{newsfeed} for each origin $i\in\mathcal{N}$. After that, we can map each vector $\mathbf{p}_i$ to $\mathbf{q}_i$ through \eqref{wall} and finally using \eqref{eq_psiscore} we can derive each $\psi_i$.\\

As shown in \cite[Theorem 5]{psiscore}, in the homogeneous case when all users in the network have the same posting and re-posting activity, i.e. $\forall n \in \mathcal{N}, \;\lambda^{(n)}=\lambda$ and $\mu^{(n)}=\mu$, it holds $\bm{\psi} = \bm{\pi}$ where $\bm{\pi}$ is the PageRank vector with a damping factor of $\alpha = \frac{\mu}{\lambda + \mu}$.\\

\textbf{Problem Statement:} The current computation of the $\psi$-score is very slow (compared e.g. to PageRank), as it requires to solve $N$ systems of $N$ equations each, where $N$ is the number of users. This is especially problematic when we want to calculate the $\psi$-score in real networks with very large number of users.

Hence, given a directed social graph $\mathcal{G}=(\mathcal{N}, \mathcal{E})$ where each node has information about its activity of posting and sharing, we aim for an algorithm that computes the $\psi$-score for all nodes in the graph as fast as PageRank, which is the classically used alternative.

\section{Proposed Method}
\label{sec:III}

In this section, we present our main contribution: a distributed algorithm for the fast approximation of the $\psi$-score, to any given requirement of error-tolerance. To attain this goal, we first derive a new expression for the $\psi$-score that  reduces its computational complexity from solving $N$ linear systems of size $N$, to just one linear system of size $N$ summarizing all users. Then, we show that power-iteration rules can be used to approximate the solution of such linear system, something that can furthermore profit by distributed computations. This way, we obtain an algorithm for the derivation of the $\psi$-score that matches the empirical complexity of PageRank, while we benefit by the expressiveness of the score due to the incorporation of richer information. 

\subsection{Expressing with a single linear system}

For the sake of notation clarity, we will express the $\psi$-score of the network in terms of the matrices $\mathbf{P, Q, B, D}$, which are defined in Table \ref{tab:notations}. This matrix notation allows us to express the vector of $\psi$-scores as
\begin{eqnarray}
\label{psimat}
\bm{\psi}^T = \frac{1}{N} \mathbf{1}^T  \mathbf{Q}
\end{eqnarray}

Thus, the problem of computing the $\psi$-score amounts to computing a matrix $\mathbf{Q}$, whose columns require knowledge of the vectors $\mathbf{p}_i$. Finding each $\mathbf{p}_i$ amounts to solving a linear system as shown in Eq. \eqref{newsfeed}. Note in \eqref{newsfeed}-\eqref{wall} that irrespective of the user origin $i$, all equations use the same matrices $\mathbf{A}$ and $\mathbf{C}$. The solution of each system can be expressed as (see \cite[Lemma 2]{psiscore}):
\begin{equation}
    \label{nf_i_solution}
    \mathbf{p}_i = (\mathbf{I} - \mathbf{A})^{-1} \mathbf{b}_i = \sum\limits_{t=0}^\infty{\mathbf{A}^t} \mathbf{b}_i
\end{equation}
The linearity of \eqref{nf_i_solution} implies that the solution for all users can be written in a single matrix form as
\begin{equation}
    \label{nf_solution}
    \mathbf{P} = (\mathbf{I} - \mathbf{A})^{-1} \mathbf{B} = \sum\limits_{t=0}^\infty{\mathbf{A}^t} \mathbf{B}
\end{equation}

\noindent Then, by leveraging \eqref{wall} in matrix form and the series development in  (\ref{nf_solution}), we can further develop Eq. (\ref{psimat}) as follows:
\begin{align*}
    \bm{\psi}^T &= \frac{1}{N} \mathbf{1}^T \mathbf{Q}\\
         &= \frac{1}{N} \mathbf{1}^T (\mathbf{CP} + \mathbf{D})\\
         &= \frac{1}{N} \mathbf{1}^T \left[ \C (\I-\A)^{-1}\B + \D\right]\\
         &= \frac{1}{N} \mathbf{1}^T \left[\C \left(\sum\limits_{t=0}^{\infty}\A^t \right) \B + \D\right]\\
         &= \frac{1}{N} \left[\mathbf{1}^T\C \left(\sum\limits_{t=0}^{\infty}\A^t \right) \B + \mathbf{1}^T\D\right]\\
         &= \frac{1}{N} \left[\left(\sum\limits_{t=0}^{\infty}\mathbf{1}^T\C\A^t \right) \B + \mathbf{1}^T\D\right]\\
\end{align*}

Now, if we let $\mathbf{c}^T := \mathbf{1}^T \mathbf{C}$ and $\mathbf{d}^T := \mathbf{1}^T \mathbf{D}$, we have that $\mathbf{c}$ and $\mathbf{d}$ are vectors in $\mathbb{R}_+^N$ (the diagonals of the matrices $\C$ and $\D$). These derivations allow us to state our main theoretical result, which is the following expression for the $\psi$-score:
\begin{eqnarray}
    \label{ultimsol}
    \bm{\psi}^T = \frac{1}{N} \left[\left(\sum\limits_{t=0}^{\infty}\mathbf{c}^T\A^t \right) \B + \mathbf{d}^T\right]
\end{eqnarray}
Eq. \eqref{ultimsol} indicates that it is not necessary to solve $N$ linear systems of size $N$ in order to compute the $\psi$-score. Instead, it suffices to solve one linear system of size $N$ which is expressed as an infinite sum of vector matrix products, and can even be calculated distributedly. The drawback is that we lose with this new expression the calculation of intermediate detailed influence quantities $\mathbf{p}_i$ and $\mathbf{q}_i$. These are not necessary to be explicitly known for the calculation of the $\psi$-score as \eqref{ultimsol} shows, but they do contain valuable information that could be useful for certain practical applications. Accelerating the detailed calculation of all by-products of the $\psi$-score is an important challenge for future work.

\subsection{Convergence of the sum}

In this subsection, we demonstrate that the infinite sum in Eq. \eqref{ultimsol} can be evaluated by power-iteration rules. The sum always converges (for the specific expression of $\mathbf{A}$) and can thus be truncated to a finite number of terms to find an approximation of the $\psi$-score. To show this, let us define:
\begin{eqnarray}
    \label{infsum}
    \mathbf{s}_t^T = \sum\limits_{\tau=0}^{t}\mathbf{c}^T\A^{\tau}, & & \mathbf{s} = \lim_{t\rightarrow\infty} \mathbf{s}_t,
\end{eqnarray}
where $T$ denotes (matrix or vector) transpose, $t$ is the iteration index and $\tau$ the running summation index. We stress that Eq. \eqref{infsum} converges if the spectral radius bound of $\mathbf{A}$ is in the open unit interval. This convergence is guaranteed by Lemmas 1 and 2 of \cite{psiscore} which use the fact that $\mathbf{A}$ is a sub-stochastic matrix. 

\subsection{Truncation of the sum}
We now proceed to study the approximation error after truncating the sum to the first $t$ terms. For this, we express the truncation of the sum in the following recursive way:
\begin{equation}
    \label{power_psi}
    \mathbf{s}_t^T = \mathbf{s}_{t-1}^T\mathbf{A} + \mathbf{c}^T
\end{equation}
where $\mathbf{s}_0 = \mathbf{c}$. This allows us to define the following gap parameter
\begin{equation}
    \label{eps_t}
    \varepsilon_t = \norm{\mathbf{s}_{t}^T - \mathbf{s}_{t-1}^T}
\end{equation}
which is useful for the termination rule of the algorithm. We do not specify which norm to use in the gap formula because all the work done in this part is valid with any norm. Nevertheless, in our implementation, we choose to use the $L^1$-norm as it is the common way to deal with series approximation with power-iterations \cite{pagerank}. We stop the recursion if the trajectory towards the fixed point between two successive iterations does not change, i.e. when the following condition is satisfied:
\begin{equation}
    \label{condition}
    \varepsilon_t \leq \varepsilon
\end{equation}

An important question to address is therefore how the tolerance set to terminate the series summation $\mathbf{s}_t$ in \eqref{infsum} affects the approximation of the $\psi$-score in \eqref{ultimsol}. To address this point, let us define $\bm{\psi}_{t}$ as the approximation of $\bm{\psi}$ by replacing \eqref{power_psi} into \eqref{ultimsol} using $t$ terms. Then, we can define the gap between two approximations of the $\psi$-score with different truncations as: 
\begin{equation}
    \label{delta_t}
    \delta_t = \norm{\bm{\psi}_{t}^T - \bm{\psi}_{t-1}^T}
\end{equation}
Our goal is thus to express $\delta_t$ as a function of $\varepsilon_t$ or to find an upper-bound relation between them to not only set a convergence condition for the vector $\mathbf{s}_t$ but more importantly for the vector $\bm{\psi}_t$. We derive this relationship by expressing the difference $\bm{\psi}_t - \bm{\psi}_{t-1}$ in terms of $\mathbf{s}_{t-1}$ and $\mathbf{s}_{t}$ as follows:
\begin{align} \label{psi_diff}
    \nonumber \bm{\psi}_t^T - \bm{\psi}_{t-1}^T &= \frac{1}{N}(\mathbf{s}_t^T \B + \mathbf{d}^T) - \frac{1}{N} (\mathbf{s}_{t-1}^T \B + \mathbf{d}^T)\\
    &= \frac{1}{N} (\mathbf{s}_{t}^T - \mathbf{s}_{t-1}^T) \B
\end{align}

Eq. \eqref{psi_diff} then allows to see that:
\begin{align}
    \nonumber \delta_t &= \frac{1}{N} \norm{(\mathbf{s}_t^T - \mathbf{s}_{t-1}^T) \B}\\
    \nonumber &\leq \frac{1}{N} \norm{\mathbf{s}_t^T - \mathbf{s}_{t-1}^T} \norm{\B}\\
    \delta_t &\leq \frac{\varepsilon_t \norm{\B}}{N},
\end{align}
With this upper bound, we can thus select a termination tolerance $\varepsilon$ that guarantees that the $\psi$-score trajectory did not change more than $\delta$. Namely, if we terminate the algorithm by satisfying the condition $\varepsilon_t \norm{\B} \leq \varepsilon$, then we are certain that $\delta_t$ is lower than $\frac{\varepsilon}{N}$. We coin our algorithm for the fast approximation of the $\psi$-score as \texttt{Power-$\psi$}. It is summarized in Algorithm \ref{power_psi_algo}.

\begin{algorithm}[t]
\caption{\texttt{Power-$\psi$:} Power iteration based algorithm for the $\psi$-score vector.}\label{power_psi_algo}
\SetAlgoLined
\SetKwData{Left}{left}\SetKwData{This}{this}\SetKwData{Up}{up}\SetKwFunction{Union}{Union}\SetKwFunction{FindCompress}{FindCompress}
\SetKwInOut{Input}{input}\SetKwInOut{Output}{output}
\Input{$N$ number of users, $N\times N$ matrices $\A$ and $\B$, two vectors $\mathbf{c}$ and $\mathbf{d}$, $\mathbf{s}$-tolerance $\varepsilon$}
\Output{vector $\bm{\psi}$ with the $\psi$-score of all users}
$\mathbf{s} \leftarrow \mathbf{c} $\;
$B\_norm \leftarrow \norm{\B}$\;
$t \leftarrow 0$\;
$gap \leftarrow 1$\;
\While{
$(gap > \varepsilon)$
}
{
    $\mathbf{s}_{old} \leftarrow \mathbf{s}$\;
    $\mathbf{s}^T \leftarrow \mathbf{s}_{old}^T \A + \mathbf{c}$\;
    $gap \leftarrow B\_norm \norm{\mathbf{s}_{old}-\mathbf{s}}$\;
    $t \leftarrow t + 1$\;
}
$\bm{\psi}^T \leftarrow \frac{1}{N} \left(\mathbf{s}^T \B + \mathbf{d}^T \right)$\;
return $\bm{\psi}$\;
\end{algorithm}

\subsection{Relationship between $\psi$-score and PageRank}
\label{pr_vs_psi}

For the sake of completeness, in this subsection we discuss the relation between equation \eqref{power_psi} in the homogeneous activity case and PageRank \cite{pagerank}. Then, we highlight the difference between our proposed method and the standard power-iteration algorithm for PageRank.

Let us define $\bm{\pi}$ as the PageRank vector and $\mathbf{W}=\mathbf{D}_{out}^{-1}\mathbf{L}$ the random walk transition matrix where $\mathbf{L}$ is the adjacency matrix and $\mathbf{D}_{out}$ is the diagonal matrix in which each diagonal entry $(i,i)$ is the out-degree of node $i$. It was shown in \cite[Theorem 5]{psiscore} that in the homogeneous activity case (where $\lambda^{(i)}= \lambda$ and $\mu^{(i)}=\mu$, $\forall i$) the $\psi$-score coincides with PageRank. This is, $\bm{\psi}=\bm{\pi}$ with $\alpha=\frac{\mu}{\lambda+\mu}$ being PageRank's damping factor.

In the case of homogeneous activity, the vectors and matrices involved in the $\psi$-score expression of Eq. \eqref{ultimsol} take the following form:

\begin{align*}
    \A &= \alpha \mathbf{W}\\
    \B &= (1-\alpha) \mathbf{W}\\
    \mathbf{c} &= \alpha \mathbf{1}\\
    \mathbf{d} &= (1-\alpha) \mathbf{1}.
\end{align*}
Then, after substitution in Eq. \eqref{power_psi}, it becomes:
\begin{equation}
    \label{psi_pagerank}
    \mathbf{s}_t^T = \alpha \mathbf{s}_{t-1}^T \mathbf{W} + \alpha \mathbf{1}^T
\end{equation}

By substituting the above expression in \eqref{ultimsol} using the homogeneous values for $\mathbf{B}$ and $\mathbf{d}$ we get
\begin{eqnarray}
\bm{\psi}_t^T & = & \frac{1}{N}\left(\mathbf{s}_t^T(1-\alpha)\mathbf{W}+(1-\alpha)\mathbf{1}^T\right)\nonumber\\
& = &  \frac{1}{N}\left((\alpha \mathbf{s}_{t-1}^T \mathbf{W} + \alpha \mathbf{1}^T)(1-\alpha)\mathbf{W}+(1-\alpha)\mathbf{1}^T\right)\nonumber\\ 
& = & \alpha\left(\frac{1}{N}( \mathbf{s}_{t-1}^T(1-\alpha) \mathbf{W} +(1-\alpha) \mathbf{1}^T)\right)\mathbf{W}+\nonumber\\
& + & \frac{(1-\alpha)}{N}\mathbf{1}^T\nonumber\\
& = & \alpha\bm{\psi}_{t-1}^T\mathbf{W}+\frac{(1-\alpha)}{N}\mathbf{1}^T
\end{eqnarray}
For comparison, the above iteration actually evolves as the power-method for PageRank
\begin{equation}
    \label{pagerank_power}
    \bm{\pi}_t^T = \alpha \bm{\pi}_{t-1}^T \mathbf{W} + \frac{1-\alpha}{N} \mathbf{1}^T.
\end{equation}

Notice that the proposed \texttt{Power-$\psi$} algorithm in the general non-homogeneous case uses power-iteration to first approximate the series $\mathbf{s}$ up to some truncation step $t$ based on the tolerance required; after convergence it  multiplies by $\mathbf{B}$ and adds $\mathbf{d}$ at the very end, to avoid unnecessary matrix vector multiplications. This is the main difference in implementation compared to the power-iteration with PageRank. As shown above, both methods essentially converge to the same value in the homogeneous case, but the critical detail is the tolerance, which in the case of $\psi$-score is on the $\mathbf{s}$-vector, whereas in the case of PageRank is on the PageRank-vector itself.

\section{The \texttt{psi-score} Python package}
\label{sec:IV}

The \textit{psi-score} project\footnote{\url{https://github.com/NouamaneA/psi-score}} is a software library that we have developed in Python. It is open source and licensed under the terms of the MIT license, allowing free usability.

This package is a practical tool for the community to easily use the $\psi$-score metric in Network Science projects without the need to develop each algorithm individually.
Both methods mentioned in this work are accessible.

Inspired by \textit{scikit-network} \cite{sknet} and \textit{scikit-learn} \cite{sklearn_api}, the project has a similar Application Programming Interface (API) in order to be user-friendly.

\section{Numerical Evaluation}
\label{sec:V}

In this section, we aim to evaluate the performance of our Power-$\psi$ algorithm (Algorithm \ref{power_psi_algo}). We employ the following two metrics to assess complexity: $(i)$ the number of matrix-vector multiplications to reach a targeted tolerance; and $(ii)$ the actual runtime of the algorithms in a machine with the following characteristics: 256GB of RAM, 4 CPUs (with 6 cores) at 3GHz. The former allows to assess and compare the complexity of distributed algorithms irrespective of the technology where the algorithms are implemented. The latter allows to get insights on the scalability of the algorithms in a real-world use case.

We run three series of experiments. The first one aims to validate the ability of Power-$\psi$ to approximate the true $\psi$-score by comparing the approximation error \texttt{Power-$\psi$} does with the current state-of-the-art alternative. The second one aims to compare the performance of \texttt{Power-$\psi$} with \texttt{Power-NF} and \texttt{PageRank} in terms of required number of matrix-vector multiplications. The third experiment shows that the proposed \texttt{Power-$\psi$} scales to large graphs. It aims to compare the computation time for the three mentioned methods.

Each experiment has been run twice: $(i)$ one time to compute the $\psi$-score with heterogeneous activity for the nodes, i.e. $\lambda^{(i)}\neq \lambda^{(j)}$ and $\mu^{(i)}\neq \mu^{(j)}$ for at least one pair of nodes $i,j$ (in practice, the activity frequencies have been chosen uniformly at random within the interval $(0,1)$); $(ii)$ the second time we run the experiments to compute the $\psi$-score algorithm with homogeneous activity $\lambda^{(n)}=\lambda=0.15$ and $\mu^{(n)}=\mu=0.85$ for all $n$ (case that reduces to PageRank as discussed in Section \ref{pr_vs_psi}) and compare it to the standard power-iteration for PageRank with $\alpha=\frac{\mu}{\lambda + \mu}=0.85$ for consistency in our comparisons.

In the experiments we vary the quantity $\varepsilon$ related to tolerance. Note here that each one of the algorithms (\texttt{Power-$\psi$}, \texttt{Power-NF} and \texttt{PageRank}'s power-method) have a different notion of tolerance, depending on the criterion of convergence each algorithm applies. Specifically, for \texttt{PageRank} we refer to $\pi$-tolerance  $\norm{\bm{\pi}_t^T-\bm{\pi}_{t-1}^T}\leq \varepsilon$ because the convergence is based on the change in PageRank values; for \texttt{Power-NF} we refer to $p$-tolerance $\norm{\mathbf{p}_i(t) - \mathbf{p}_i(t-1)}\leq \varepsilon$ because the convergence is based on the change in the news-feed $p$-values; for \texttt{Power-$\psi$} we refer to $s$-tolerance $\norm{\mathbf{s}_t^T - \mathbf{s}_{t-1}^T}$ because as shown above the convergence is based on the change in the series $s$-values. Note in the case of \texttt{Power-NF} and  \texttt{Power-$\psi$} further processing needs to be done after convergence in order to eventually derive the resulting $\psi$-scores. Hence, when we set some value $\varepsilon$ for the x-tolerance in a specific method, we are interested to evaluate the resulting relative error in the result $\bm{\psi}_{\varepsilon}$ of this method compared to the true value $\bm{\psi}_{true}$, defined as follows
\begin{equation}
    \label{error}
    error = \frac{\norm{\bm{\psi}_{true} - \bm{\psi}_{\varepsilon}}_2}{\norm{\bm{\psi}_{true}}_2}
\end{equation}

Table \ref{tab:datasets} lists the datasets used for our experiment, which are available in the Konect repository \cite{konect}. Twitter and Facebook graphs, where users have posting and re-posting activity, are natural application scenarios for our method. The two citation datasets (DBLP and HepPh arXiv) are also relevant application cases for our algorithm. This is because paper publications per author can be seen as equivalent to posting activity, and paper citations by author can be seen as equivalent to reposting activity. In all datasets, only the graph topology is available, thus user activity has been generated either at random for $(i)$ or chosen for $(ii)$ in order to have $\frac{\mu}{\lambda + \mu}=0.85$.

In order to be able to assess the error with respect to a true $\psi$-score, for experiments 1 and 2 we focus entirely on the DBLP citation network \cite{DBLP:conf/spire/Ley02} that has a moderate size and it is thus realistic to exactly solve the linear system. Experiment 3 addresses the remainder of datasets from Table~\ref{tab:datasets}.

Our implementation is available on GitHub\footnote{\url{https://github.com/NouamaneA/fast-algorithm-psiscore}} as well as the code for numerical evaluation.

\begin{table}[htbp]
    \centering
    \caption{Datasets used for evaluation}
    \begin{tabular}{|c|c|c|c|c|}
        \hline
        \textbf{Dataset name} & \textbf{Type} & \textbf{\#Nodes} & \textbf{\#Edges} & \textbf{References} \\ \hline
        \texttt{DBLP} & Citation Network & 12 591 & 49 743 & \cite{konect,DBLP:conf/spire/Ley02}\\
        \texttt{Twitter} & Social Network & 465 017 & 834 797 & \cite{konect, twitter_dataset}\\
        \texttt{Facebook} & Social Network & 63 731 & 817 035 & \cite{konect, fb_dataset}\\
        \texttt{HepPh arXiv} & Citation Network & 34 546 & 421 578 & \cite{konect, hepph_dataset}\\ \hline
    \end{tabular}
    \label{tab:datasets}
\end{table}

\subsection{Experiment 1}

In the first experiment, we assess the approximation error made with \texttt{Power-$\psi$}. We compare it with the approximation errors made with the other methods \texttt{Power-NF} and \texttt{PageRank}'s power-method. This experiment seeks to show that our proposed algorithm can approximate the $\psi$-score vector (also the PageRank vector in the homogeneous case) at least as well as existing methods. 

We compute the true $\psi$-score vector: once for the heterogeneous case $(i)$ and a second time for the homogeneous case. Then, we run each algorithm with several targeted tolerance from $10^{-9}$ to $10^{-1}$. Finally we calculate the approximation error with equation \eqref{error} for each realization of each method.

\begin{figure}[htbp]
    \centering
    \includegraphics[width=\linewidth]{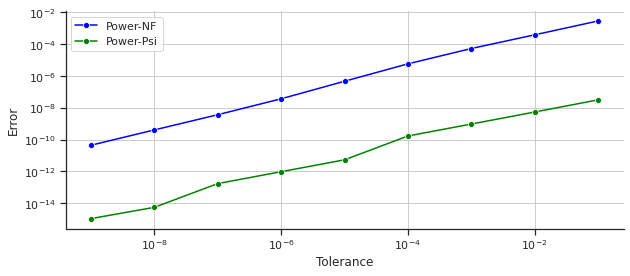}
    \caption{Experiment 1 $(i)$ with \texttt{DBLP}: Precision assessment of the proposed algorithm compared to the state-of-the-art method. Heterogeneous case where users do not necessarily have the same activity.}
    \label{fig:exp1_psi}
\end{figure}

\begin{figure}[htbp]
    \centering
    \includegraphics[width=\linewidth]{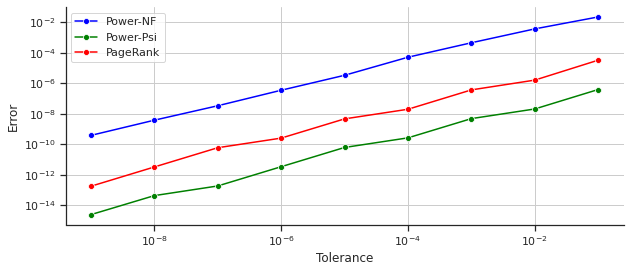}
    \caption{Experiment 1 $(ii)$ with \texttt{DBLP}: Precision assessment of the proposed algorithm compared to the state-of-the-art method. Homogeneous case where all users have the same activity and $\psi$-score=PageRank.}
    \label{fig:exp1_pr}
\end{figure}

The results are shown for heterogeneous activity $(i)$ in Fig.~\ref{fig:exp1_psi} and for homogeneous activity $(ii)$ in Fig.~\ref{fig:exp1_pr}. We fix the same level of $\mathbf{p}$- and $\mathbf{s}$-tolerance for both algorithms \texttt{Power-NF} and \texttt{Power-$\psi$} as well as PageRank power iteration and derive the error from \eqref{error}. We see in Fig.~\ref{fig:exp1_psi} and Fig.~\ref{fig:exp1_pr}, for the same targeted tolerance, that the error made with \texttt{Power-$\psi$} method is lower than the approximation error made with \texttt{Power-NF} and \texttt{PageRank}, which validates the convergence of our method to the true value.

\subsection{Experiment 2}

The second experiment evaluates the number of matrix-vector multiplications that are required to compute the $\psi$-score and compare the methods with the same precision.

\begin{figure}[htbp]
    \centering
    \includegraphics[width=\linewidth]{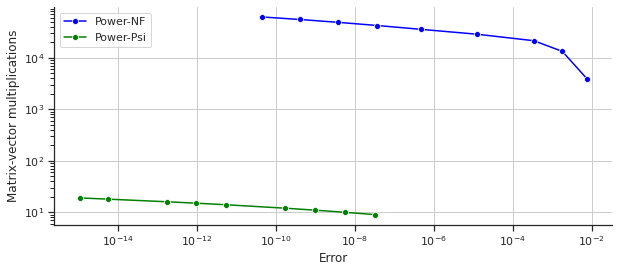}
    \caption{Experiment 2 $(i)$ with \texttt{DBLP}: Comparison of the proposed algorithm with the state-of-the-art method in terms of number of matrix-vector multiplications. Heterogeneous case where users do not necessarily have the same activity.}
    \label{fig:exp2_psi}
\end{figure}

\begin{figure}[htbp]
    \centering
    \includegraphics[width=\linewidth]{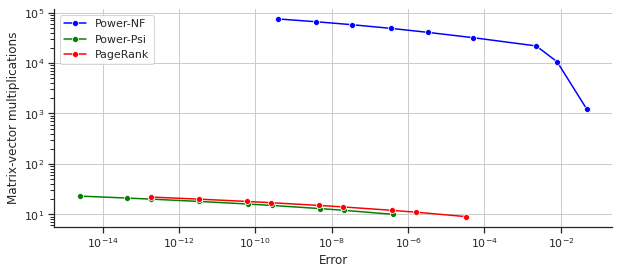}
    \caption{Experiment 2 $(ii)$ with \texttt{DBLP}: Comparison of the proposed algorithm with the state-of-the-art method and PageRank's power-method in terms of number of matrix-vector multiplications. Homogeneous case where all users have the same activity and $\psi$-score=PageRank.}
    \label{fig:exp2_pr}
\end{figure}

We run the proposed algorithm and the state-of-the-art algorithm presented in \cite{psiscore} with different tolerance parameters from $10^{-9}$ from $10^{-1}$. We measure the number of matrix-vector multiplications done for each method as well as the relative error calculated with \eqref{error} in order to compare the algorithms within a given precision. Figure \ref{fig:exp2_psi} shows the result in the case where we compute the $\psi$-score with $(i)$ heterogeneous activity and Figure \ref{fig:exp2_pr} shows the result when computing $\psi$-score in the specific case of $(ii)$ homogeneous activity that gives the PageRank vector.

We see clearly that the proposed algorithm \texttt{Power-$\psi$} outperforms the old one \texttt{Power-NF} by several orders of magnitude. We can also observe that the difference in the number of matrix-vector multiplications compared to PageRank is very small, which renders the two methods computationally equivalent, at least for the power-iteration method.

\subsection{Experiment 3}

In the third experiment, we seek to demonstrate that our proposed algorithm scales to real-world networks with several hundreds of thousands of edges, at least as well as PageRank does. For this purpose, we chose one targeted tolerance (for space reasons) and we measured the computation time after running each method. Table~\ref{tab:time_i} shows the results we obtained in the case where we compute the $\psi$-score with heterogeneous activity $(i)$ and Table~\ref{tab:time_ii} shows the results in the case of homogeneity $(ii)$ when the $\psi$-score vector equals the PageRank vector.

Here again, the proposed \texttt{Power-$\psi$} outperforms by far the state-of-the-art \texttt{Power-NF}. We can also conclude that the proposed algorithm is close to PageRank's power-method in terms of computation time. In some cases, \texttt{Power-$\psi$} takes longer but that is because the algorithm needs to perform other operations after the convergence of the approximated series.

\begin{table}[htbp]
    \centering
    \caption{Experiment 3 $(i)$: Computation time evaluation with $\varepsilon$ set to $10^{-9}$. Heterogeneous case where users do not necessarily have the same activity.}
    \begin{tabular}{|c|c|c|}
        \hline
        Dataset & Power-NF \cite{psiscore} & Power-$\psi$\\ \hline
        \texttt{DBLP} & 17.805 sec & 0.029 sec\\
        \texttt{Facebook}  & 1764.226 sec & 0.307 sec\\
        \texttt{Twitter}  & 14526.039 sec & 0.634 sec\\
        \texttt{HepPh} & 272.358 sec & 0.622 sec\\ \hline
    \end{tabular}
    \label{tab:time_i}
\end{table}
\begin{table}[htbp]
    \centering
    \caption{Experiment 3 $(ii)$: Computation time evaluation with $\varepsilon$ set to $10^{-9}$. Homogeneous case where all users have the same activity and $\psi$-score=PageRank.}
    \begin{tabular}{|c|c|c|c|}
        \hline
        Dataset & PageRank \cite{pagerank} & Power-NF \cite{psiscore} & Power-$\psi$\\ \hline
        \texttt{DBLP} & 0.023 sec & 20.775 sec & 0.034 sec\\
        \texttt{Facebook}  & 0.308 sec & 2253.302 sec & 0.454 sec\\
        \texttt{Twitter} & 0.584 sec & 17411.146 sec & 0.806 sec\\
        \texttt{HepPh} & 0.361 sec & 360.769 sec & 0.908 sec\\ \hline
    \end{tabular}
    \label{tab:time_ii}
\end{table}

\section{Conclusion and Future Work}
\label{sec:VI}

We have proposed a new method to compute the $\psi$-score, which is influence of users in a social network (or in any network with nodes having a posting a sharing activity). We have shown that the proposed algorithm converges faster than the state-of-the-art method. The algorithm enables scalability for real-world datasets for the $\psi$-score. 

We have also provided an open source software library that implements the existing algorithm for the $\psi$-score calculation as well as the novel approach.

The proposed method gives directly the influence of every user over the entire network. But in some cases, one might need to use the influence of a user on a specific user, which is giver by the OSP model but skipped by the proposed \texttt{Power-$\psi$} to go faster to the point. In the future, we are interested in further study on the $\psi$-score in order to explore the possibility to get these other information potentially valuable for some applications.

\section*{Acknowledgements}

The work of N.A. is funded directly by Sorbonne University, CNRS-LIP6 laboratory by an internal project. The research of A.G. is supported by the French National Agency of Research (ANR) through the FairEngine project under Grant ANR-19-CE25-0011. E.B. is funded by the French National Agency of Research (ANR) under the FiT LabCom grant.

\bibliographystyle{IEEEtran}
\bibliography{references}

\begin{thebibliography}{10}
\providecommand{\url}[1]{#1}
\csname url@samestyle\endcsname
\providecommand{\newblock}{\relax}
\providecommand{\bibinfo}[2]{#2}
\providecommand{\BIBentrySTDinterwordspacing}{\spaceskip=0pt\relax}
\providecommand{\BIBentryALTinterwordstretchfactor}{4}
\providecommand{\BIBentryALTinterwordspacing}{\spaceskip=\fontdimen2\font plus
\BIBentryALTinterwordstretchfactor\fontdimen3\font minus
  \fontdimen4\font\relax}
\providecommand{\BIBforeignlanguage}[2]{{%
\expandafter\ifx\csname l@#1\endcsname\relax
\typeout{** WARNING: IEEEtran.bst: No hyphenation pattern has been}%
\typeout{** loaded for the language `#1'. Using the pattern for}%
\typeout{** the default language instead.}%
\else
\language=\csname l@#1\endcsname
\fi
#2}}
\providecommand{\BIBdecl}{\relax}
\BIBdecl

\bibitem{dkempe}
\BIBentryALTinterwordspacing
D.~Kempe, J.~Kleinberg, and E.~Tardos, ``Maximizing the spread of influence
  through a social network,'' in \emph{Proceedings of the Ninth ACM SIGKDD
  International Conference on Knowledge Discovery and Data Mining}, ser. KDD
  '03.\hskip 1em plus 0.5em minus 0.4em\relax New York, NY, USA: Association
  for Computing Machinery, 2003, p. 137–146. [Online]. Available:
  \url{https://doi.org/10.1145/956750.956769}
\BIBentrySTDinterwordspacing

\bibitem{alvim2021multi}
M.~S. Alvim, B.~Amorim, S.~Knight, S.~Quintero, and F.~Valencia, ``A
  multi-agent model for polarization under confirmation bias in social
  networks,'' in \emph{International Conference on Formal Techniques for
  Distributed Objects, Components, and Systems}.\hskip 1em plus 0.5em minus
  0.4em\relax Springer, 2021, pp. 22--41.

\bibitem{bok2021hot}
K.~Bok, Y.~Noh, J.~Lim, and J.~Yoo, ``Hot topic prediction considering
  influence and expertise in social media,'' \emph{Electronic Commerce
  Research}, vol.~21, no.~3, pp. 671--687, 2021.

\bibitem{MustoLGS21}
\BIBentryALTinterwordspacing
C.~Musto, P.~Lops, M.~de~Gemmis, and G.~Semeraro, ``Context-aware graph-based
  recommendations exploiting personalized pagerank,'' \emph{Knowl. Based
  Syst.}, vol. 216, p. 106806, 2021. [Online]. Available:
  \url{https://doi.org/10.1016/j.knosys.2021.106806}
\BIBentrySTDinterwordspacing

\bibitem{cabannes2021overcoming}
V.~Cabannes, L.~Pillaud-Vivien, F.~Bach, and A.~Rudi, ``Overcoming the curse of
  dimensionality with laplacian regularization in semi-supervised learning,''
  \emph{Advances in Neural Information Processing Systems}, vol.~34, 2021.

\bibitem{diffusion_networks}
\BIBentryALTinterwordspacing
S.~Goel, D.~J. Watts, and D.~G. Goldstein, ``The structure of online diffusion
  networks,'' in \emph{Proceedings of the 13th ACM Conference on Electronic
  Commerce}, ser. EC '12.\hskip 1em plus 0.5em minus 0.4em\relax New York, NY,
  USA: Association for Computing Machinery, 2012, p. 623–638. [Online].
  Available: \url{https://doi.org/10.1145/2229012.2229058}
\BIBentrySTDinterwordspacing

\bibitem{centrality}
Z.~Wan, Y.~Mahajan, B.~W. Kang, T.~J. Moore, and J.-H. Cho, ``A survey on
  centrality metrics and their network resilience analysis,'' \emph{IEEE
  Access}, vol.~9, pp. 104\,773--104\,819, 2021.

\bibitem{Cha_Haddadi}
\BIBentryALTinterwordspacing
M.~Cha, H.~Haddadi, F.~Benevenuto, and K.~Gummadi, ``Measuring user influence
  in twitter: The million follower fallacy,'' \emph{Proceedings of the
  International AAAI Conference on Web and Social Media}, vol.~4, no.~1, pp.
  10--17, May 2010. [Online]. Available:
  \url{https://ojs.aaai.org/index.php/ICWSM/article/view/14033}
\BIBentrySTDinterwordspacing

\bibitem{Sala12}
\BIBentryALTinterwordspacing
C.~Wilson, A.~Sala, K.~P.~N. Puttaswamy, and B.~Y. Zhao, ``Beyond social
  graphs: User interactions in online social networks and their implications,''
  \emph{ACM Trans. Web}, vol.~6, no.~4, nov 2012. [Online]. Available:
  \url{https://doi.org/10.1145/2382616.2382620}
\BIBentrySTDinterwordspacing

\bibitem{psiscore}
\BIBentryALTinterwordspacing
A.~Giovanidis, B.~Baynat, C.~Magnien, and A.~Vendeville, ``Ranking online
  social users by their influence,'' \emph{{IEEE/ACM} Trans. Netw.}, vol.~29,
  no.~5, pp. 2198--2214, 2021. [Online]. Available:
  \url{https://doi.org/10.1109/TNET.2021.3085201}
\BIBentrySTDinterwordspacing

\bibitem{social_network}
C.~Correa, T.~Crnovrsanin, and K.-L. Ma, ``Visual reasoning about social
  networks using centrality sensitivity,'' \emph{IEEE Transactions on
  Visualization and Computer Graphics}, vol.~18, no.~1, pp. 106--120, 2012.

\bibitem{transportation}
\BIBentryALTinterwordspacing
R.~Puzis, Y.~Altshuler, Y.~Elovici, S.~Bekhor, Y.~Shiftan, and A.~Pentland,
  ``Augmented betweenness centrality for environmentally aware traffic
  monitoring in transportation networks,'' \emph{J. Intell. Transp. Syst.},
  vol.~17, no.~1, pp. 91--105, 2013. [Online]. Available:
  \url{https://doi.org/10.1080/15472450.2012.716663}
\BIBentrySTDinterwordspacing

\bibitem{communication_network}
A.~Tizghadam and A.~Leon-Garcia, ``Betweenness centrality and resistance
  distance in communication networks,'' \emph{IEEE Network}, vol.~24, no.~6,
  pp. 10--16, 2010.

\bibitem{biology}
\BIBentryALTinterwordspacing
M.~Ashtiani, A.~Salehzadeh-Yazdi, Z.~Razaghi-Moghadam, H.~Hennig,
  O.~Wolkenhauer, M.~Mirzaie, and M.~Jafari, ``A systematic survey of
  centrality measures for protein-protein interaction networks,''
  \emph{bioRxiv}, 2017. [Online]. Available:
  \url{https://www.biorxiv.org/content/early/2017/10/09/149492}
\BIBentrySTDinterwordspacing

\bibitem{politics}
\BIBentryALTinterwordspacing
P.~R. Miller, P.~S. Bobkowski, D.~Maliniak, and R.~B. Rapoport, ``Talking
  politics on facebook: Network centrality and political discussion practices
  in social media,'' \emph{Political Research Quarterly}, vol.~68, no.~2, pp.
  377--391, 2015. [Online]. Available:
  \url{https://doi.org/10.1177/1065912915580135}
\BIBentrySTDinterwordspacing

\bibitem{pagerank}
L.~Page, S.~Brin, R.~Motwani, and T.~Winograd, ``The pagerank citation ranking:
  Bringing order to the web,'' 1998.

\bibitem{scalable_pagerank}
J.~Whang, A.~Lenharth, I.~Dhillon, and K.~Pingali, ``Scalable data-driven
  pagerank: Algorithms, system issues, and lessons learned,'' 08 2015, pp.
  438--450.

\bibitem{DBLP:journals/snam/BautistaL22}
\BIBentryALTinterwordspacing
E.~Bautista and M.~Latapy, ``A local updating algorithm for personalized
  pagerank via chebyshev polynomials,'' \emph{Soc. Netw. Anal. Min.}, vol.~12,
  no.~1, p.~31, 2022. [Online]. Available:
  \url{https://doi.org/10.1007/s13278-022-00860-5}
\BIBentrySTDinterwordspacing

\bibitem{sknet}
\BIBentryALTinterwordspacing
T.~Bonald, N.~de~Lara, Q.~Lutz, and B.~Charpentier, ``Scikit-network: Graph
  analysis in python,'' \emph{Journal of Machine Learning Research}, vol.~21,
  no. 185, pp. 1--6, 2020. [Online]. Available:
  \url{http://jmlr.org/papers/v21/20-412.html}
\BIBentrySTDinterwordspacing

\bibitem{sklearn_api}
L.~Buitinck, G.~Louppe, M.~Blondel, F.~Pedregosa, A.~Mueller, O.~Grisel,
  V.~Niculae, P.~Prettenhofer, A.~Gramfort, J.~Grobler, R.~Layton,
  J.~VanderPlas, A.~Joly, B.~Holt, and G.~Varoquaux, ``{API} design for machine
  learning software: experiences from the scikit-learn project,'' in \emph{ECML
  PKDD Workshop: Languages for Data Mining and Machine Learning}, 2013, pp.
  108--122.

\bibitem{konect}
\BIBentryALTinterwordspacing
J.~Kunegis, ``{KONECT} -- {The} {Koblenz} {Network} {Collection},'' in
  \emph{Proc. Int. Conf. on World Wide Web Companion}, 2013, pp. 1343--1350.
  [Online]. Available: \url{http://dl.acm.org/citation.cfm?id=2488173}
\BIBentrySTDinterwordspacing

\bibitem{DBLP:conf/spire/Ley02}
\BIBentryALTinterwordspacing
M.~Ley, ``The {DBLP} computer science bibliography: Evolution, research issues,
  perspectives,'' in \emph{String Processing and Information Retrieval, 9th
  International Symposium, {SPIRE} 2002, Lisbon, Portugal, September 11-13,
  2002, Proceedings}, ser. Lecture Notes in Computer Science, A.~H.~F. Laender
  and A.~L. Oliveira, Eds., vol. 2476.\hskip 1em plus 0.5em minus 0.4em\relax
  Springer, 2002, pp. 1--10. [Online]. Available:
  \url{https://doi.org/10.1007/3-540-45735-6\_1}
\BIBentrySTDinterwordspacing

\bibitem{twitter_dataset}
M.~{De Choudhury}, Y.~Lin, H.~Sundaram, K.~Candan, L.~Xie, and A.~Kelliher,
  ``\BIBforeignlanguage{English (US)}{How does the data sampling strategy
  impact the discovery of information diffusion in social media?}'' in
  \emph{\BIBforeignlanguage{English (US)}{ICWSM 2010 - Proceedings of the 4th
  International AAAI Conference on Weblogs and Social Media}}, ser. ICWSM 2010
  - Proceedings of the 4th International AAAI Conference on Weblogs and Social
  Media, Dec. 2010, pp. 34--41, 4th International AAAI Conference on Weblogs
  and Social Media, ICWSM 2010 ; Conference date: 23-05-2010 Through
  26-05-2010.

\bibitem{fb_dataset}
\BIBentryALTinterwordspacing
B.~Viswanath, A.~Mislove, M.~Cha, and K.~P. Gummadi, ``On the evolution of user
  interaction in facebook,'' in \emph{Proceedings of the 2nd ACM Workshop on
  Online Social Networks}, ser. WOSN '09.\hskip 1em plus 0.5em minus
  0.4em\relax New York, NY, USA: Association for Computing Machinery, 2009, p.
  37–42. [Online]. Available: \url{https://doi.org/10.1145/1592665.1592675}
\BIBentrySTDinterwordspacing

\bibitem{hepph_dataset}
\BIBentryALTinterwordspacing
J.~Leskovec, J.~Kleinberg, and C.~Faloutsos, ``Graph evolution: Densification
  and shrinking diameters,'' \emph{ACM Trans. Knowl. Discov. Data}, vol.~1,
  no.~1, p. 2–es, mar 2007. [Online]. Available:
  \url{https://doi.org/10.1145/1217299.1217301}
\BIBentrySTDinterwordspacing

\end{thebibliography}

\end{document}